\documentclass[twocolumn,showpacs,preprintnumbers,prb]{revtex4}

\begin{document}

\title[dfa]{Bare LO-Phonon Peak in THz-Emission Signals: a
Dielectric-Function Analysis}
\author{Fabr\'{\i}cio M. Souza and J. C. Egues}
\email{egues@if.sc.usp.br}
\affiliation{Departamento de F\'{\i}sica e Inform\'{a}tica, Instituto de F\'{\i}sica de S%
\~{a}o Carlos, Universidade de S\~{a}o Paulo, 13560-970 S\~{a}o Carlos, S%
\~{a}o Paulo, Brasil.}

\begin{abstract}
We present a normal-mode analysis of coupled photocarrier-phonon
dynamics in Te. We consider a dielectric function which accounts
for LO\ phonons and the electron-hole gas within the
Debye-H\"{u}ckel model and RPA. Our main finding is the existence
of a \emph{bare} LO\ phonon mode in the system even at high
carrier density. This oscillation is an unscreened $L_{-}(q) $
mode arising from ineffective screening at large wave vectors.
This mode is consistent with the bare LO-phonon peak in recent
THz-emission spectra of Te.
\end{abstract}

\date[Date: ]{\today}
\maketitle

The recent progress in ultrafast spectroscopy allows the
investigation of photocarrier-phonon coupled THz dynamics in
semiconductors \cite{dekorsy-review}. Ultrashort pulses enable the
excitation and detection of coherent infrared lattice vibrations
which is a source of THz radiation. THz emission has been measured
recently in semiconductors via a dipole antenna
with detectable frequencies up to around 5 THz \cite{telurio-prl}-\cite{dip}%
. Tellurium is a suitable semiconductor for these experiments
because it possesses infrared-active modes within the antenna
spectral range.

The physics of THz emission in Te is straightforward. A laser
pulse generates a non-equilibrium distribution of electron-hole
pairs. The absorption length at the laser energies used in the
experiments ($\sim $1.7 eV) is 40 nm \cite{note}. This small
absorption length creates a highly inhomogeneous
\emph{longitudinal} charge density and diffusion currents arise.
These currents lead to charge separation, due to the different
electron and hole mobilities, which generates a longitudinal
polarization field (Dember field). The Dember field drives LO
phonons (Te is piezoelectric) whose macroscopic oscillatory
polarization emits radiation.

The emitted spectrum from the Te crystal consists of a broad peak
at approximately 0.5 THz, followed by a dip to approximately zero
at $\nu
_{TO}= $2.6 THz and a second maximum close to $\nu _{LO}= $2.82 THz \cite%
{telurio-prb},\cite{dip}. The peak at 0.5 THz is related to the
ultrafast buildup of the Dember field \cite{telurio-prb}. The
\emph{bare} LO phonon mode is usually explained in terms of the
lateral inhomogeneity of the excitation spot. In this picture the
low-density ``wings'' of the spot gives
rise to phonon-like long wavelength ($q=0 $) modes at the bare LO frequency {%
[}see $L_{+} $ branch in Fig. 1(a){]}.

Here we use a simple normal-mode analysis to gain further insight
into the physics behind the bare LO peak observed in the emission
spectrum of Te. Interestingly enough, we find that the
longitudinal inhomogeneity of the electron-hole gas gives rise to
bare LO phonon modes even at high carrier densities. This is in
contrast with the usual picture involving lateral inhomogeneity of
$q=0 $ modes. We believe longitudinal inhomogenety ($q\neq 0 $)
leading to ineffective screening of the phonon polarization fields
by the electron-hole gas provides an alternate (or complementary)
description for the observed bare LO frequency.

\emph{Dielectric function.} In a first approximation the total
dielectric function of the electron-hole--phonon system is given
by
\begin{equation}  \label{1}
\epsilon (q,\omega )=\epsilon _{\infty }+\frac{\epsilon
_{0}-\epsilon _{\infty }}{1-(\frac{\omega }{\omega
_{TO}})^{2}}-V_{q}P(q,\omega ),
\end{equation}
with
\begin{equation}  \label{hydro}
V_{q}P(q,\omega )=\sum _{i=e,h}\frac{\omega _{pl,i}^{2}\epsilon _{\infty }}{%
\omega ^{2}-\frac{k_{B}T_{i}}{m^{*}_{i}}q^{2}},
\end{equation}
where $\omega _{pl,i}=\sqrt{\frac{4\pi n_{i}e^{2}}{\epsilon
_{\infty }m^{*}_{i}}} $ is the plasma frequency for electrons
($i=e $) or holes ($i=h
$), $m^{*}_{i} $ is the effective mass ($m^{*}_{e/h}=0.067/0.53\ m_{0} $), $%
n_{i} $ is the particle density, $e $ is the electron charge,
$\epsilon _{\infty } $ = 10.9 is the dielectric constant for
frequencies much above the reststrahl $\omega \gg \omega _{TO} $ =
$2\pi \: \nu _{TO} $, $\epsilon _{0} $ = 12.8 is the static
dielectric constant, $T_{i} $ is the carrier temperature and
$k_{B} $ is the Boltzmann constant. Apart from a $\epsilon
_{\infty } $ additive term, Eq. (\ref{hydro}) is essentially the Debye-H\"{u}%
ckel dielectric function within a hydrodynamic model \cite{rick}. Note that $%
\epsilon (q,\omega ) $ takes into account the longitudinal
inhomogeneity of the electron-hole gas through the $q $ dependence
in (\ref{hydro}). Longitudinal inhomogeneity plays a fundamental
role in the Te system because it gives rise to a driving Dember
field which excites the infrared lattice modes.

\emph{Normal modes.} The zeros of the dielectric function
(\ref{1}) define the normal-mode frequencies of the system:
$\epsilon (q,\omega
)=0\Rightarrow \omega (q,n) $, where $n $ is the carrier density (we assume $%
n=n_{e}=n_{h}) $. Figure 1(a) shows the usual dispersion relation
for $q=0 $
as a function of the carrier densities \cite{mooradian} . The branches $%
L_{+} $ and $L_{-} $ (solid lines) arise from the anti-crossing of
LO phonons and plasmons. The important point here is that the
$L_{-} $ branch goes to the screened LO frequency $\nu _{TO} $ as
$n $ increases. This is due to screening of the lattice
polarization field by the plasma.

Figure 1(b) shows the $L_{-} $ branch for $n=10^{18} $ cm$^{-3} $
as a function of $q $. For $q\rightarrow 0 $ we recover the
screened LO frequency since $n $ is high. When $q $ increases,
however, screening becomes inefficient and the $L_{-} $ branch
shifts to the bare LO frequency $\nu _{LO} $\cite{cowley}. Hence
screening ineffectiveness is a possible mechanism to generate a
bare LO peak in the spectrum of the electron-hole-phonon system.

It is instructive to compare the Debye-H\"{u}ckel and RPA
dielectric function analyses. The RPA dielectric function is given
by \cite{mahan},
\begin{equation}
V_{q}P(q,\omega )=\frac{4\pi e^{2}}{q^{2}\Omega }\sum_{i=e,h}\sum_{\mathbf{k}%
}\frac{f_{i}(\xi _{i\mathbf{k}})-f(\xi
_{i\mathbf{k}+\mathbf{q}})}{\omega -\xi _{i\mathbf{k}}+\xi
_{i\mathbf{k}+\mathbf{q}}+i\delta }.  \label{rpa}
\end{equation}%
where $\Omega $ is the sample volume, $f_{i}(\xi _{i\mathbf{k}})$
is the Fermi distribution function for electrons or holes and $\xi
_{i\mathbf{k}}$ is the free electron or hole energy. Figure 2(b)
clearly shows that RPA and Debye-H\"{u}ckel yield similar
dispersion curves for $L_{-}(q)$. This is consistent with the high
temperatures used  in the calculation ($T_{e/h}=7000/3600$ K)
\cite{note2}. Note that we are \emph{not} assuming $q\rightarrow
0$ in the RPA contribution (\ref{rpa}).

The main result of our simple normal-mode analysis -- namely, that
the longitudinal inhomogeneity of the electron-hole gas makes
screening ineffective thus allowing for \emph{bare} LO modes even
at high densities -- also follows from a more detailed numerical
investigation \cite{fab-egues1}. We have solved a set of coupled
differential equations consisting of Poisson's equation, a
driven-harmonic-oscillator equation and transport equations which
include inhomogeneity via diffusion terms. The numerical spectra
are similar to the experimental ones, i.e., they also show a broad
peak near 0.5 THz, a dip at 2.6 THz and a second peak close to 2.82 THz \cite%
{fab-egues}.

In conclusion, the dielectric function analysis presented here
reveals that non-zero wave vector modes lead to bare LO phonon
oscillations of the electron-hole-phonon system even at high
carrier density. This is due to the suppression of the screening
of the lattice polarization field by the inhomogeneous
electron-hole gas. This screening ineffectiveness provides an
alternative/complementary description for the observed bare LO
frequency in THz emission experiments.

The authors acknowledge support from FAPESP and CAPES.

\begin{figure}[tbp]
\caption{Plasmon-phonon coupled modes for Tellurium as a function
of carrier density [for $q=0$] (a) and wave vector $q$ (b). The
lower branch $L_{-}$ (a) tends to the TO frequency value for high
densities due to the efficient screening of the lattice
polarization field by the electron-hole gas. Ineffective screening
on the other hand, relevant for non-zero wave vectors $q$ (b),
makes the $L_{-}$ branch move towards the LO frequency limit for
increasing $q$'s -- even at high densities. This is in contrast
with the $q=0$ mode in (a).} \label{fig1}
\end{figure}

\end{document}